\documentclass[%
 reprint,
 amsmath,amssymb,
 aps,
]{revtex4-2}

\usepackage{graphicx}% Include figure files
\usepackage{dcolumn}% Align table columns on decimal point
\usepackage{bm}% bold math
\usepackage{physics}% For Bra-Ket notation 
\usepackage{caption}
\usepackage{subcaption}
\usepackage{xcolor}
\usepackage{hyperref}% add hypertext capabilities
\begin{document}

\preprint{APS/123-QED}

\title{Improving robustness of quantum feedback control with reinforcement learning}% Force line breaks with \\

\author{Manuel Guatto}
\affiliation{Dipartimento di Ingegneria dell’Informazione, Universit\'a degli Studi di Padova, via Gradenigo 6B, 35131 Padova, Italy}
\affiliation{Forschungszentrum J\"ulich, Institute of Quantum Control (PGI-8), D-52425 J\"ulich, Germany}
\author{Gian Antonio Susto}
\affiliation{Dipartimento di Ingegneria dell’Informazione, Universit\'a degli Studi di Padova,
via Gradenigo 6B, 35131 Padova, Italy}
\affiliation{Human Inspired Technology Research Center, Universit\'a degli Studi di Padova, via Luigi Luzzatti, 4, 35121 Padova PD, Italy}
\author{Francesco Ticozzi}
\affiliation{Dipartimento di Ingegneria dell’Informazione, Universit\'a degli Studi di Padova,
via Gradenigo 6B, 35131 Padova, Italy}
\affiliation{Department of Physics and Astronomy, Dartmouth College, 6127 Wilder Laboratory, Hanover, NH 03755, USA}

 

% \collaboration{CLEO Collaboration}%\noaffiliation

\date{%
    \today
}% It is always \today, today,

\begin{abstract}
Obtaining reliable state preparation protocols is a key step towards practical  implementation of many quantum technologies, and one of the main tasks in quantum control.
In this work, different reinforcement learning approaches are used to derive a feedback law for state preparation of a desired state in a target system. In particular, we focus on the robustness of the obtained strategies with respect to different types and amount of noise. Comparing the results indicates that the learned controls are more robust to unmodeled perturbations with respect to simple feedback strategy based on optimized population transfer, and that training on simulated nominal model retain the same advantages displayed by controllers trained on real data. The possibility of effective off-line training of robust controllers promises significant advantages towards practical implementation.
\end{abstract}

 %Use showkeys class option if keyword
\maketitle

%\tableofcontents

\section{Introduction}

The development of reliable control tools for state preparation and engineering of desired dynamics in target quantum systems is a key instrumental step toward quantum information processing on a scale that is suitable for real world applications \cite{altafini,dalessandro,koch2022quantum,dong}.  The main tool used by control systems developers to design robust control laws that are able to adapt to unforeseen conditions and steer classical systems towards the target behavior is feedback: measuring some key quantity in real time and using the acquired information to allows, among other things, for disturbance rejection and noise mitigation \cite{classicalfeedback}. 
In the quantum domain, measurements and feedback become nontrivial, as they introduce probabilistic evolution and back action in the picture: models and methods for quantum feedback have been extensively studied \cite{belavkin1,belavkin2,wiseman,wisemanbook,barchielli,mirrahimi,transfer,bouten,gough,zhang2017quantum,nurdin2009coherent,ticozzi2009analysis,doherty}, as well as successfully demonstrated experimentally \cite{haroche}.
Nonetheless, the robustness advantage of feedback methods have been demonstrated analytically to hold for quantum implementations only in particular cases \cite{ticozzi2004robust,van2007filtering,petersen2012robust,liang2021robust,ge2012quantum}, mostly with respect to uncertainty on the initial state or delays in the feedback loop, with some general results on model perturbations just starting to be derived. \cite{robustness}.

In this work, we investigate the potential role of data-based learning methods in deriving feedback strategy and their robustness with respect to unknown model perturbations. Reinforcement learning (RL) has already been considered as a control design tool for quantum systems in \cite{Sivak_2022,Porotti_2022,DBLP:journals/corr/BarryBA14,stable-baselines3,Bukov_2018,Zhang_2019,Guo_2021,Mackeprang_2020,Baba_2023,preti2023hybrid,chen2013fidelity,Porotti_2019,Haug_2020,Bukov_2018_Floquet,Borah_2021,PhysRevX.8.031084,PhysRevA.94.042122,Nature551.579,2211.03464,2210.16715,PhysRevLett.128.060502,PhysRevResearch.2.012011,PhysRevA.97.052346,PhysRevLett.124.050501,tseng2019reinforcement,chen2022reinforcement,zhou2022reinforcement,song2022reinforcement,luo2022reinforcement}, both towards control and error-correction tasks. With respect to the existing results, we focus on {\em comparing} different techniques in order to investigate:
\begin{itemize}
    \item The role of the a-priori knowledge on the model, by using different training scenarios for the controller;
    \item The robustness of RL methods, also compared to basic controllers, by introducing un-modeled noise in the evolution;
    \item The role of the measurement accuracy in addressing the control task under;
    \item The viability of of RL based on nominal models, rather than real data.
\end{itemize}
We consider the last point to be particularly interesting, and often overlooked in the literature, as the overhead of running an experiment long enough to grant convergence for the RL method could be a reason alone to discourage the use of these methods in practical scenarios.

The analysis we propose is based on numeric simulations on a testbed system, focusing on feedback  state-preparation problems. Intuitively, the problem of interest for this work can be formulated as follows: given a quantum system in contact with a Markovian environment and subjected to repeated measurements, design a feedback control law that implements unitary control action dependent on the measurement outcomes and the system knowledge, aimed to minimize the distance between a desired target state $\rho_{\text{target}}$ and the actual state $\rho(T)$ at the end of a finite time horizon $[0, T]$.  

The results of our analysis, reported in detail Section \ref{sec:conclusions}, indicate that model-based RL feedback laws are indeed a natural candidate when modeling errors and noise are expected, and that an accurate measurement is crucial to obtain high-fidelity.  

The rest of the paper is organized as follows: Section \ref{sec:model} introduces the general model for our discrete-time feedback system, and Section \ref{sec:testbed} specializes it to the system that will be simulated. A short review of the key ideas behind the RL methods we employ is provided in Section \ref{sec:RL}, while Section \ref{sec:scenarios} details the basic control strategy as well as three learning scenarios we consider; two learning scenarios employ a quantum description of the system and the same RL method, based on a Proximal Policy Optimization (PPO) algorithm, while the third does not rely on any a-priori knowledge on the model and builds the control law based only on {\em classical} data, i.e. the output of the measurement. The numerical experiments we conducted are reported in Section \ref{sec:numerical} and the main findings summarized in Section \ref{sec:conclusions}.

\section{\label{sec:model}  Discrete-time feedback state-preparation with noise and generalized measurements}

In this section, we present in detail the mathematical model for the feedback state preparation problem described above. We define the various elements of the model, introduce the notation, and illustrate the interplay between quantum operations, measurements, and control parameters. In this paper only finite-dimensional systems are considered.

For open quantum systems undergoing Markovian evolutions, the transition of the system's state in Schroedinger's picture is associated to a linear, Completely-Positive Trace-Preserving (CPTP) map \cite{nielsen-chuang}. These maps admits a Kraus representation:
\begin{equation}
{\cal E}(\rho_t) = \sum_k E_k \rho_t E_k^{\dagger},
\end{equation}
where the operators $E_k$ satisfy $\sum_k  E_k^{\dagger}E_k=I.$ In our setting, quantum operations will be used to describe both the noise and the control actions. 

We assume our system to undergo a time-homogeneous Markovian noisy dynamics associated to a CPTP map $\mathcal{N}_\alpha,$ with the parameter $\alpha$ to weigh the amount of noise injected in the system. Such map is represented by the ensemble of Kraus operators $\{N_k^\alpha\}$.

The noise action is followed by a generalized measurement with a finite set of outcomes $l=1,\ldots , m,$ associated to a set of measurement operators $M_l^\epsilon$, where $\epsilon$ is a parameter associated to how informative the measurements are. In particular $\epsilon = 0$ will correspond to projective measurements. The operators are such that $\sum_l M_l^{\epsilon\dag} M_l^\epsilon=I.$ The probability $p(l_t)$ for a specific outcome $l$ at time $t$ is computed as
\begin{equation}
p(l_t) = \mathrm{tr}(M_{l}^{\epsilon\dag} M_{l}^\epsilon \rho_t').
\end{equation}
Once the measurement outcome $l_t$ is obtained, the post-measurement state $\rho_{t+1 | l_t}$ is updated as follows:
\begin{equation}
\mathcal{M}_{l_t,\epsilon} (\rho_t) = \rho_{t+1 | l_t} = \frac{M_{l_t}^\epsilon\rho_t M_{l_t}^{\epsilon\dagger}}{\mathrm{tr}(M_{l_t}^{\epsilon\dagger} M_{l_t}^\epsilon \rho_t)}.
\end{equation}

The measurements are followed by a unitary control action, conditioned on the last measurement outcome. The unitary control is assumed to take the form $U_\beta = e^{-iH_c (\beta)}$, with $H_c$ a control Hamiltonian and $\beta$ the control parameter. In the following we will consider  feedback laws of the form $\beta_t=\phi(\rho_0,\vec l_{t-1}),$ where $\vec l_{t-1}$ represents the sequence of outcomes up to time $t-1$ and the functional $\phi$ can be determined either analytically or via  reinforcement learning. The unitary control super-operator then becomes $\mathcal{U}_\beta$:
\begin{equation}
\mathcal{U}_\beta(\rho) = U_\beta \rho U_\beta^\dagger.
\end{equation}
The evolution of the system is then obtained by iterating these steps. From a physical viewpoint, we consider the actual time-scales of the measurement and control processes to be faster than the noise ones, so we can neglect the noise while measurements and control are acting. The dynamical evolution, conditional on a sequence of measurements outcomes $\vec l_{t}$ is obtained as the composition of $\mathcal{N}_\alpha$, $\mathcal{U}_{\phi(\rho_0,\vec l_{t-1})}$, and ${\cal M}_{l_t,\epsilon}$:
\begin{equation}\label{eq:dynamics}
\begin{cases}
\rho({t+1}) =  {\cal  M}_{l_t,\epsilon} \circ \mathcal{U}_{\phi(\rho_0,\vec l_{t-1})}\circ\mathcal{N}_\alpha[\rho(t)]\\
\rho(0)=\rho_0
\end{cases},
\end{equation}
with $\rho_0$ the initial condition for the dynamics.
While the concatenated dynamics is not a CPTP map, as it is not linear due to the conditioning, its expectation over the measurements outcomes is such, and takes the form:
\begin{equation}\label{eq:avgdynamics}
\rho(t+1) = \sum_{k,l}M_l^\epsilon U_{\phi(\rho_0,l_{t-1})}N_k^\alpha\rho(t) N_k^{\alpha\dag}U^\dag_{\phi(\rho_0,l_{t-1})}M_l^{\epsilon\dag}.
\end{equation}

The design of stabilizing control laws $\phi$ is aimed to obtain convergence of the quantum state $\rho(t)$ towards the target state $\rho_{\text{target}}$ \cite{bolognani} independently of the initial condition, namely $\forall \rho(0)$ we want the dynamics above to ensure:
\begin{equation}
    \lim_{t\to\infty} \rho(t) = \rho_{\text{target}},
\end{equation}
where $\rho_{\text{target}}(t)$ is the desired target state at time step $t$. 

In what follows, we consider instead the related problem of {\em state preparation in finite time}, where the control task is, for some fixed time $T$, to:
\begin{equation}\label{eq:stateprep}
    \textrm{minimize}_{\phi} \;d(\rho(T)-\rho_\text{target}),
\end{equation}
where $d$ is a suitable norm or pseudo-distance. A stabilizing law on the infinite control horizon then represents an approximate solution for the finite-time state preparation problem, with its accuracy improving for larger $T$.

In the subsequent sections, we propose different ways to build feedback strategies under different assumptions regarding the available information on the system, and compare their robustness and performance towards state preparation in finite time.

\section{A testbed system} \label{sec:testbed} 
For our numerical analysis, we consider a 3-level quantum system system inspired by that of \cite{Porotti_2022}.  Define the Hilbert space of interest as $\mathcal{H} = \mathbb{C}^{3}$, and the basis vector as:
\begin{equation}
    \ket{0} = \begin{pmatrix}
    1 \\
    0\\
    0
    \end{pmatrix}, \ket{1} = \begin{pmatrix}
    0\\
    1\\
    0
    \end{pmatrix}, \ket{2} = \begin{pmatrix}
    0\\
    0\\
    1
    \end{pmatrix}.
\end{equation}

\subsection{Noise}
Two key noise models, namely the depolarizing noise and the random permutation channel, are considered in the simulations presented in the next section. 

The {\em depolarizing channel} is a fundamental noise model that describes the effects of ``isotropic'' random errors and disturbances in quantum systems. It can be defined through its action on a quantum state $\rho$ as follows:

\begin{equation}
\mathcal{N}_\alpha (\rho) = \alpha \frac{I}{3} + (1-\alpha) \rho.
\end{equation}
$\mathcal{N}_\alpha(\rho)$ represents the quantum state after the application of the depolarizing noise and $\alpha \in [0, 1]$ is a parameter that quantifies the strength of the noise.\\

The \emph{amplitude damping channel} for a qutrit system is a non-unital noise which reduces the energy of the system due to an interaction with the environment. The amplitude damping channel is described by its Kraus representation $\{\ket{0}, \ket{1}, \ket{2}\}$ defined as follows:
\begin{equation}
    \begin{aligned}
    N_0 = \begin{pmatrix}
        1 & 0 & 0\\
        0 & \sqrt{1-\gamma_1} & 0\\
        0 & 0 & \sqrt{1-\gamma_2-\gamma_3}
        \end{pmatrix},\\
    N_{01} = \begin{pmatrix}
        0 & \sqrt{\gamma_1} & 0\\
        0 & 0 & 0\\
        0 & 0 & 0
        \end{pmatrix}, N_{12} = \begin{pmatrix}
        0 & 0 & 0\\
        0 & 0 & \sqrt{\gamma_2}\\
        0 & 0 & 0
        \end{pmatrix},\\ 
    N_{03} = \begin{pmatrix}
        0 & 0 & \sqrt{\gamma_3}\\
        0 & 0 & 0\\
        0 & 0 & 0
        \end{pmatrix},
    \end{aligned}
\end{equation}
with the following constraint:
\begin{equation*}
    \begin{cases}
    &0 \leq \gamma_i \leq 1 \quad \forall i \in\{1,2,3\}\\
    &\gamma_2+\gamma_3 \leq1
    \end{cases}\quad.
\end{equation*}
In particular, in the simulated system we will set the $\gamma_i$ parameters in function of a single parameter $\alpha$: 
\begin{equation*}
    \begin{cases}
        &\gamma_1 = 0\\
        &\gamma_2+\gamma_3 = \alpha\\
        &\gamma_2 = \gamma_3
    \end{cases}\quad.
\end{equation*}

The {\em random permutation channel} or random qutrit flip channel, is a type of quantum noise that captures randomized cycling between quantum states. To describe this channel for our qutrit system, we utilize a set of permutation matrices:

\begin{equation}
    \begin{aligned}
    N_0^\alpha &= \sqrt{1-\frac{2 \alpha}{3}}  I,\\
    N_1^\alpha &= \sqrt{\frac{\alpha}{3}} \begin{pmatrix}
    0 & 0 & 1\\
    1 & 0 & 0\\
    0 & 1 & 0
    \end{pmatrix},\\
    N_2^\alpha &= \sqrt{\frac{\alpha}{3}} \begin{pmatrix}
    0 & 1 & 0\\
    0 & 0 & 1\\
    1 & 0 & 0
    \end{pmatrix},
    \end{aligned}
\end{equation}
with the corresponding CPTP map defined as follows:
\begin{equation}
    \mathcal{N}_\alpha(\rho) = \sum_k N_k^\alpha \rho N_k^{\alpha^\dagger}
\end{equation}

Similar to the depolarizing channel noise, the parameter $\alpha$ plays a crucial role in characterizing the random qutrit flip channel. It defines the probability of the qutrit system undergoing a flip operation. A larger value of $\alpha$ indicates a higher likelihood of state flips, while a smaller value corresponds to a reduced probability of flipping.

\subsection{Measurements}
The measurement process we consider is associated to a set of operators, denoted as {$M_0^\epsilon, M_1^\epsilon, M_2^\epsilon$}, which  satisfy the completeness constraint:
\begin{equation}
I = \sum_{i = 0} M_i^{\epsilon\dagger} M_i^\epsilon.
\end{equation}
In our simulated quantum system, we consider ``imprecise'' version of projective measurement, which provide useful yet incomplete information about the basis state in which the system is in. The $\epsilon$ parameter limits the amount of information provided by the measurements.

To represent this operator, we define its Kraus representation as follows:

\begin{equation}
\begin{aligned}
M_0^\epsilon &= \begin{pmatrix}
\sqrt{1-2\epsilon} & 0 & 0\\
0 & \sqrt{\epsilon} & 0\\
0 & 0 & \sqrt{\epsilon}
\end{pmatrix}, \\
M_1^\epsilon &= \begin{pmatrix}
\sqrt{\epsilon} & 0 & 0\\
0 & \sqrt{1-2\epsilon} & 0\\
0 & 0 & \sqrt{\epsilon}
\end{pmatrix}, \\
M_2^\epsilon &= \begin{pmatrix}
\sqrt{\epsilon} & 0 & 0\\
0 & \sqrt{\epsilon} & 0\\
0 & 0 & \sqrt{1-2\epsilon}
\end{pmatrix}.
\end{aligned}
\end{equation}
% from which we can define the conditional map:
% \begin{equation}
%     \mathcal{M}_{l
% _t,\epsilon}(\rho_t) = \frac{M_{l_t,\epsilon} \rho_t M_{l_t,\epsilon}^{\dagger}}{\mathrm{tr}(M_{l_t,\epsilon}^{\dagger} M_{l_t,\epsilon} \rho_t)}.
% \end{equation}
Notice that: (1) each basis state is invariant for conditioning on {\em any } output of the measurement; (2) the variable $\epsilon$ introduces an error in detecting the correct basis state, which is correctly represented by the outcome with probability $1-2\epsilon.$
\subsection{Unitary Control and State-Preparation Problem}
Recall from the previous section that the control action is of the form

\begin{equation}
U_{\beta}(t) = e^{-iH_c(\beta_t)},
\end{equation}

where we recall that $\beta_t$ represents the time-dependent control parameter, acting on a time-scale much faster than the noise one. Hence, its action can be considered decoupled from the rest of the dynamics and generating impulsive unitary control actions. For our model, we consider:
\begin{equation}
H_c(\beta_t) = i[\beta_t (a - a^\dagger)],
\end{equation}
where $\beta_t$ is a real-valued function of time with codomain between -1 and 1, and the operator $a$ is given by:
\begin{equation}
a = \begin{pmatrix}
0 & 1 & 0\\
0 & 0 & 1\\
0 & 0 & 0
\end{pmatrix}.
\end{equation}
To this unitary matrix corresponds the control superoperator $
\mathcal{U}_{\beta}(\rho) = U_{\beta} \rho U_{\beta}^\dagger$ as defined before.

The target for designing the causal feedback law $\beta_t=\phi(\rho_0,\vec l_{t-1})$ is to obtain a state as close as possible to $\rho_{\text {target}}=\ket{2}\bra{2}$ at the end of the control period $T$. Notice that since the outcomes $\{l_t\}$ form a discrete-time stochastic process, both $\rho$ and $\beta_t$ become stochastic. 
More precisely, we aim to solve problem \eqref{eq:stateprep} with pseudo distance $d(\rho,\rho_{\text target}) = 1 - F({\rho}, \rho_{target}),$ where the fidelity is defined as: 
\begin{equation}
     F({\rho}, \rho_{target}) = {\text tr}\left[\sqrt{\sqrt{\rho}\rho_{target}\sqrt{{\rho}}}\right]^2.
     \label{eq: Fidelity}
\end{equation}

\section{Reinforcement Learning: A brief Introduction}\label{sec:RL} 
\subsection{Essential Elements}

Reinforcement Learning (RL) can be considered as a subfield of machine learning that focuses on training intelligent agents to make sequential decisions through interaction with their training environment. Unlike supervised learning, where models are provided with labeled data, or unsupervised learning, which seeks to uncover hidden patterns in data, RL is focused on the learning by an \emph{agent} of optimal \emph{policies} in an \emph{environment} that provides feedback in the form of rewards \cite{Sutton1998}. Optimality, in the context of RL is associated with long-term objectives, aiming at maximizing a (weighted) sum of rewards in subsequent step of a trajectory.

In the RL field, the state space, denoted as $S$, represents the set of all possible states in which an agent can be. Formally, $S = \{s_1, s_2, ..., s_n\}$, where $s_i$ is an individual state \footnote{We here use the standard RL nomenclature, but note that this is not necessarily a (quantum) state. Similarly, in a RL context, the  environment is actually the system of interest, see Subsection \ref{sec:RLtranslated} for more details.}. The state space can be discrete or continuous, and it captures all relevant information about the environment's current configuration. To determine the action that transitions the agent from $s_t$, the state at time $t$, to the next state $s_{t+1}$, the agent relies on a defined policy. Formally, this latter can be defined as a map that determines the probability distribution of selecting actions $(a)$ in a given state $(s)$. It is represented as $\pi(a|s)$. States in RL are considered Markovian, ie. the current state contains all the relevant information for the agent and its dynamic, making the information about the past states irrelevant.

Moreover, we define a reward function, denoted as $R$, this quantifies the immediate consequence of the agent's action. Formally, $R(s, a)$ maps a state-action pair $(s, a)$ to a real number, representing the reward obtained when applying action $a$ from state $s_{t+1}$. Formally, the agent's goal is to maximize the expected cumulative reward, often expressed as the expected return, defined as:
$\mathbb{E}\left[\sum_{t=0}^{\infty} \gamma^t R(s_t, a_t, s_{t+1})\right]$,
where $\gamma \in [0,1]$ represents a discount factor.

\begin{figure}[h]
    \centering
    \includegraphics[width = 8.6cm, height = 4.5cm]{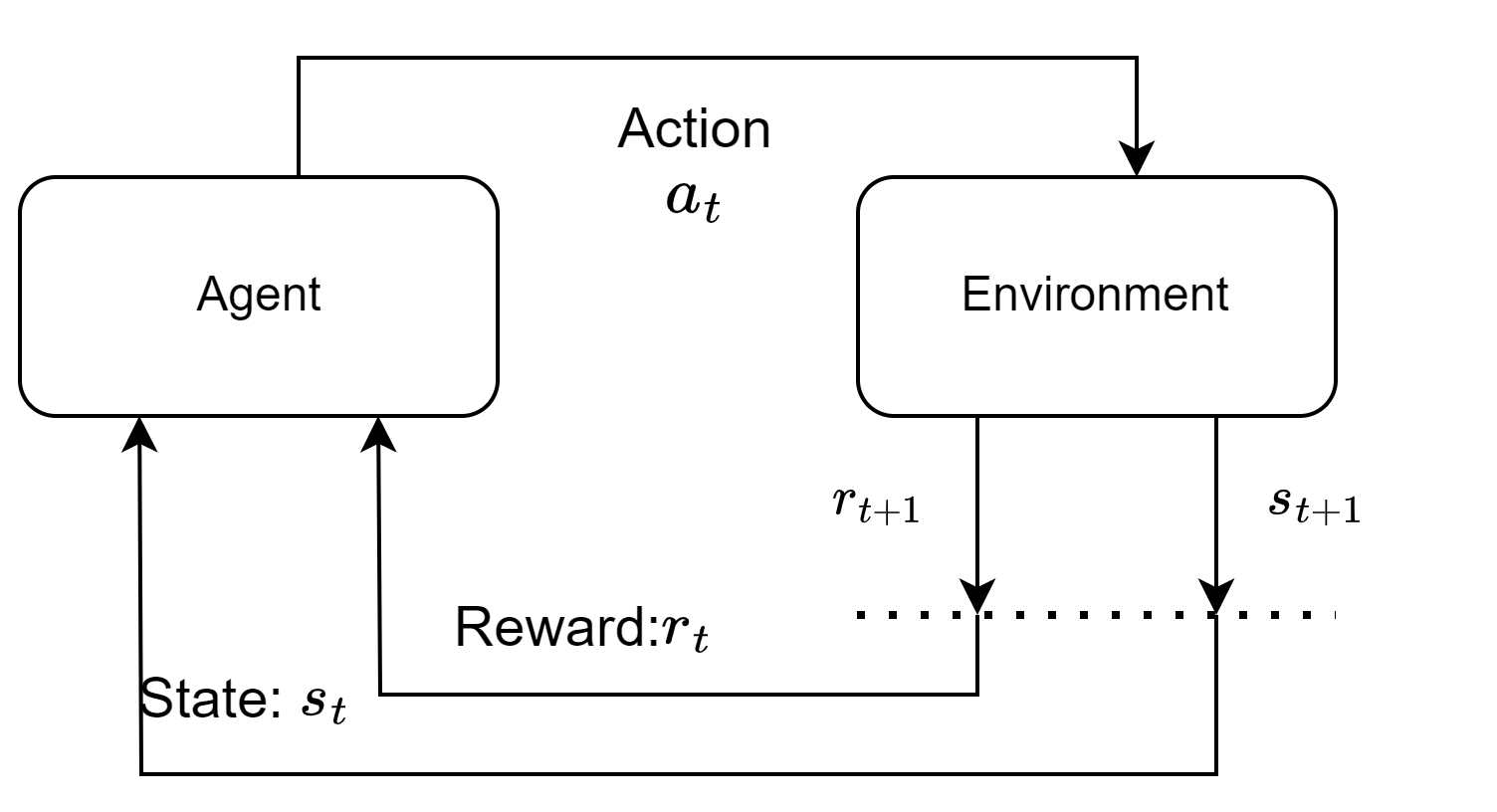}
    \caption{Reinforcement Learning interaction scheme}
    \label{fig:RL_interaction}
\end{figure}

\subsection{Policy Gradient approach}
In the realm of Reinforcement Learning, Q-learning and Policy Gradient (PG) methods represent two distinct approaches.
Q-learning estimates the value of state-action pairs and works well in discrete action spaces with fully observable environments. In contrast, PG directly optimizes policies, making it more effective for continuous action spaces and non-fully observable environments, where discretization and complete state information can be challenging for Q-learning. In this work we concentrate on PG methods: this section provides a brief introduction about the foundational principles of PG approaches.

We recall that a policy $\pi(a|s)$ determines the probability of selecting action $a$ given the current state $s$ of the RL environment. In the PG framework we assume that the each policy is associated to a set of parameters $\theta.$ 

As previously highlighted, the objective of RL lies in finding an optimal policy that maximize the expected return; in the PG framework this goal can be achieved by iteratively updating the $\theta$ parameters via the policy gradient RL update rule:

\begin{equation}
\delta\theta_j = \eta \frac{\partial\mathbb{E}[R]}{\partial\theta_j} = \eta \mathbb [ R \sum_t \frac{\partial}{\partial\theta_j} \ln \pi_\theta(a_t | s_t) ] .  
\label{eq: policy_gradient}
\end{equation}

Here, $\eta$ denotes the learning rate parameter which is a real value parameter on which the converging proprieties of the RL algorithm depend (usually it takes values between $10^{-3}$ and $10^{-5}$) and $\mathbb{E}[\cdot]$ represents the expectation value computed over all possible rewards.
These fundamental elements form the core policy gradient approach. In practical applications, enhancements and refinements of the above equation are often employed. 

Furthermore, Equation (\ref{eq: policy_gradient}) serves as the standard recipe for policy based RL in fully observed environments. This approach can be extended to accommodate partially observed environments, where the policy relies solely on observations rather than the complete state information. These observations offer only partial insights into the actual state of the environment, introducing additional complexities in the RL framework.

In our work, we choose to adopt the Stable Baselines3 implementation \cite{stable-baselines3} in which a neural network is used to compute the $\pi_\theta$ policy, with the multidimensional parameter ($\theta$) encompassing all the network's weights and biases. The neural network takes the current state ($s_t$) as an input vector and produces the action probabilities ($\pi_\theta$) as its output.

%%%%%%%%%%%%%%%%%%%%%
\subsection{RL for feedback quantum control}\label{sec:RLtranslated}
\iffalse
{\bf FT: This section is too long and wordy, and does not clearly convey the key messages. We need (in my opinion, maybe even as a bullet list) to say:

- we consider a quantum feedback control problem;

- who is the ``environment'' (quantum system, possibly undergoing noisy evolution, plus measurement apparatus. In some cases, the environment will also include a model of the evolution that will provide an estimated state [filtering equation]);

- what is the task of the agent (given the measurement outcome, and when available the estimated state, find the control parameter $\beta$ that yields best results for the state preparation problem at hand;

- who is the reward function.

The figure is nice but the caption needs to be revisited.
}
\fi

\begin{figure}[h]
    \centering
    \includegraphics[width = 8.5cm]{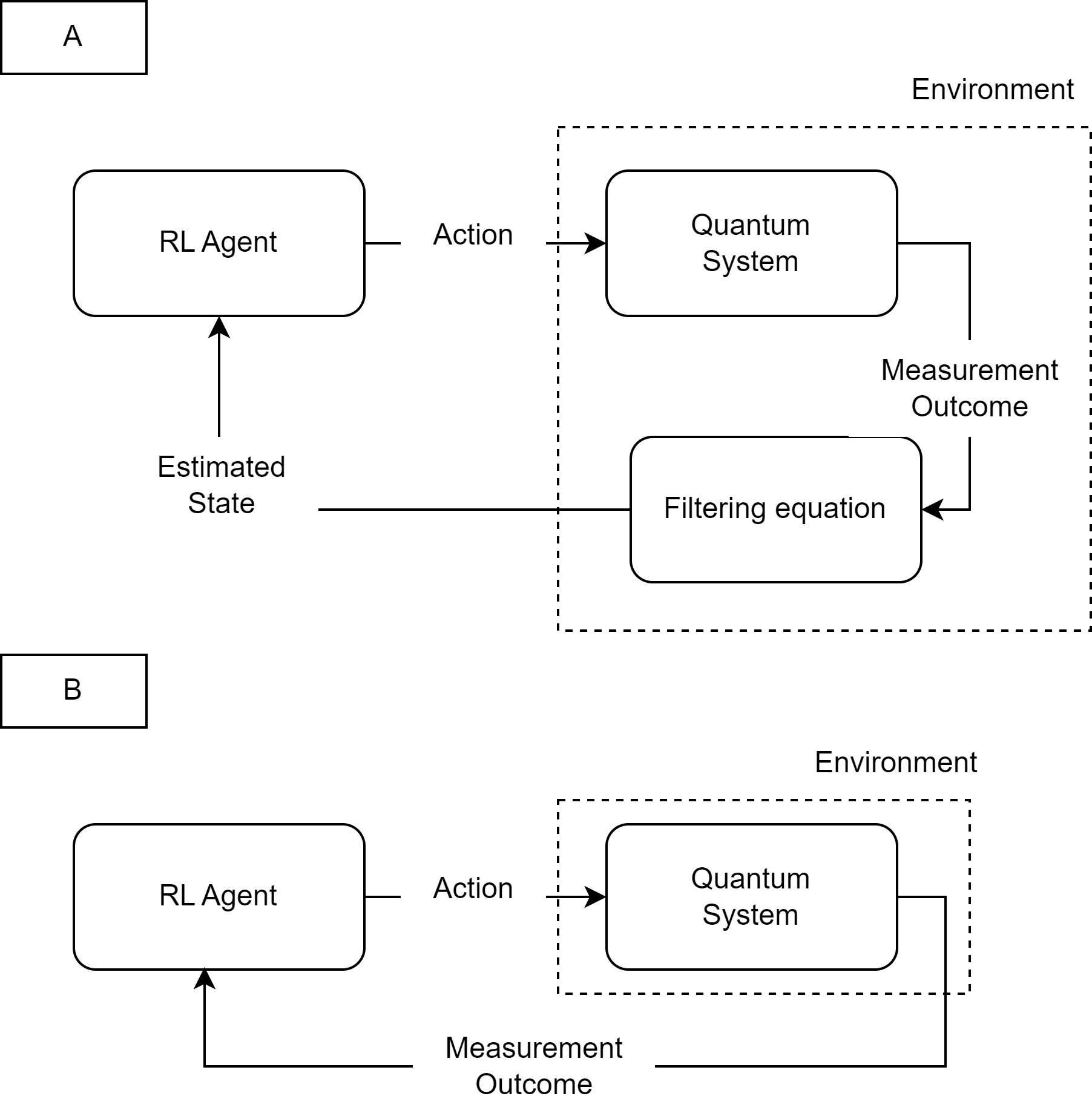}
    \caption{(A). illustrates the first feedback scenario in which the environment (square outlined with dashed lines) outputs the estimated state (B) illustrates the second feedback scenario in which the environment (square outlined with dashed lines) provides to the agent just the outcomes of the measurements performed on the quantum system}
    \label{fig:RL_Feedback}
\end{figure}

This section aims to connect the general RL framework recalled above to the specifics of the problem at hand. An in-depth discussion of the details will be provided in section \ref{sec:scenarios}.
 In a quantum feedback control problem quantum states and their dynamics are manipulated using control operations from some available set, while adapting the strategies depending on the results of quantum measurements. %The key components include the quantum system itself, quantum states described by density operators, control operations for external manipulation, probabilistic quantum measurements, and feedback strategies to adjust control based on measurement outcomes. 
Our approach consists in training a RL agent able to learn adaptive strategies through trial and error, iteratively refining its control policies based on the outcomes of quantum measurements or on an estimated state of the quantum system. The key elements of the RL setting are described in the following:

{\bf RL Environment.} The RL environment in this context is associated to a quantum system undergoing potential noisy evolution, coupled with a measurement apparatus. The best available description of the environment is quantum, where the state is described by density operators and may be subject to probabilistic changes as described in section \ref{sec:model}. The environment takes as input the control parameter $\beta$ and outputs the outcome of the measurement or, when available, the updated estimation of the state computed through a filtering equation (Fig. \ref{fig:RL_Feedback}). The selection of the environment output depends on the training model.

{\bf Agent's task.} In the context of the state preparation problem, the objective is to identify the optimal control law that determines $\beta$ based on the measurement outcome and, when accessible, on the estimated state, to drive the quantum state as close as possible to a target state. 
% This task is inherent to a RL problem, where the agent after receiving a representation of the state, engages in feedback control. Indeed the agent receives the outcome measurement or the estimated state as input and outputs as action the chosen control parameter. The primary objective therefore is to optimize the control parameter $\beta$, thereby achieving optimal results in the state preparation process. 
To do so is necessary to optimize the policy $\pi_{\theta}$ that dictates the selection of $\beta$. We want to remark that the representation of the state plays a key role in the optimization of the policy, as important as the design of the reward function.

{\bf Reward function.} The choice of the reward function is a key element towards the optimization of the agent's policy. Our work addresses two main scenarios: in the first the RL agent is supplied with an estimated density operator, while in the second the agent receives only measurement outcomes. In the former case, having access to the quantum system's density operator, we naturally opt for fidelity as the reward function, quantifying the proximity between the estimated state and the target state. In the latter scenario, characterized by a less informative environment, our strategy involves assigning a positive reward when the measurement outcome aligns with our target state and a negative reward otherwise.\\

% It is noteworthy that during the training phase of the RL algorithm, the feedback phase and the policy optimization operate synergistically. In contrast, during the validation phase, only the feedback control loop is active.

\section{Control and Learning Scenarios} \label{sec:scenarios} 
\subsection{Nominal and Filtering Dynamics} Before delving deeper into the control scenarios and in particular RL scenarios we need to define two additional states and their evolution: the \emph{nominal state} and the \emph{filtered state}.
The need for considering such states emerges because, while our RL agent is assumed to have full information about some key quantities entering the model (depending on the particular setup, these might include the initial state of the system ($\rho_0, \rho_t$), the control parameters $\beta_t$, and the measurement operator and outcomes $M_{l,\epsilon}$ and $l_t$), we suppose that it does not account for the noise present in the actual system dynamics. This will allow us to test the robustness to these strategies derived under ideal condition to the introduction of noise. 

In the light of this,  we define the nominal state as the solution to the following {\em nominal dynamics}:
\begin{equation}\label{eq:nominal}
    \begin{cases}
        \bar{\rho}(t+1) = \mathcal{M}^{\epsilon}_{\bar{l}(t)} \circ \mathcal{U}_{\beta_t}(\bar{\rho}(t))\\
        \bar{\rho}(0) = \rho_{0}
    \end{cases} ,
\end{equation}
where $\beta_t$ represents the actual control input fed to the system, and $\bar{l}(t)$ indicates the measurement outcome without noise in the system dynamics, with associated probabilities computed as:
\begin{equation}
    p(\bar{l}(t)) = tr(M_{l_t}^\dagger M_{l_t}(\bar{\rho}(t))).
\label{eq: Nominal_state_dist}
\end{equation}

Next, we define the evolution for the filtered state  $\hat{\rho}(t),$ where we consider the real measurement outcomes, but evolve the state using only the nominal, noiseless dynamics. This filtered state is then the solution of the {\em filtering dynamics}:
\begin{equation} \label{eq:filtered}
    \begin{cases}
        \hat{\rho}(t+1) = \mathcal{U}_{\beta}\mathcal{M}^{\epsilon}_{l(t)} (\hat{\rho}(t))\\
        \hat{\rho}(0) = \rho_{0}
    \end{cases} .
\end{equation}
It is worth highlighting that the difference with respect to the previous equation lays in the different origin and distribution of the outcomes: for the filtered state we consider the {\em actual} outcomes $l(t),$ with associated  distribution depending on the presence of noise and the evolution of the true state of the system $\rho(t),$ which evolves as in \eqref{eq:dynamics}:
\begin{equation}
    p(l(t)) = tr(M_{l_t}^\dagger M_{l_t}\mathcal{N}_\alpha({\rho}(t))).
    \label{eq: Filtered_distr}
\end{equation}

\subsection{Basic controller}
One possible control policy to stabilize the system described in the previous section is the use of a simple feedback law derived from the nominal system with perfect (projective) measurements, that is when $\alpha=\epsilon=0$.  The main idea is to select the control parameter according to the outcome of the most recent measurement, maximizing the probability of transition towards the target. To do so we call $\beta_0$ and $\beta_1$ the control parameter to be applied when the outcomes are respectively is $l = 0$ and $l=1$, while we pose $\beta_2=0$ since it corresponds, at least in this setting, to the target. These two fixed parameters are computed solving the following optimization problem:
\begin{equation}
\begin{split}
    \beta_0 &= \arg\max_{\beta} tr(\ket{2}\bra{2}\mathcal{U}_{\beta}(\ket{0}\bra{0}))\\
    & =  \arg\max_{\beta} tr(\bra{2}\mathcal{U}_{\beta}(\ket{0}\bra{0})\ket{2}),
\end{split}
\end{equation}
\begin{equation}
\begin{split}
    \beta_1 &= \arg\max_{\beta} tr(\ket{2}\bra{2}\mathcal{U}_{\beta}(\ket{1}\bra{1}))\\
    & =  \arg\max_{\beta} tr(\bra{2}\mathcal{U}_{\beta}(\ket{1}\bra{1})\ket{2}).
\end{split}
\end{equation}

\begin{figure}[h]
    \centering
    \includegraphics[width = 8.0cm]{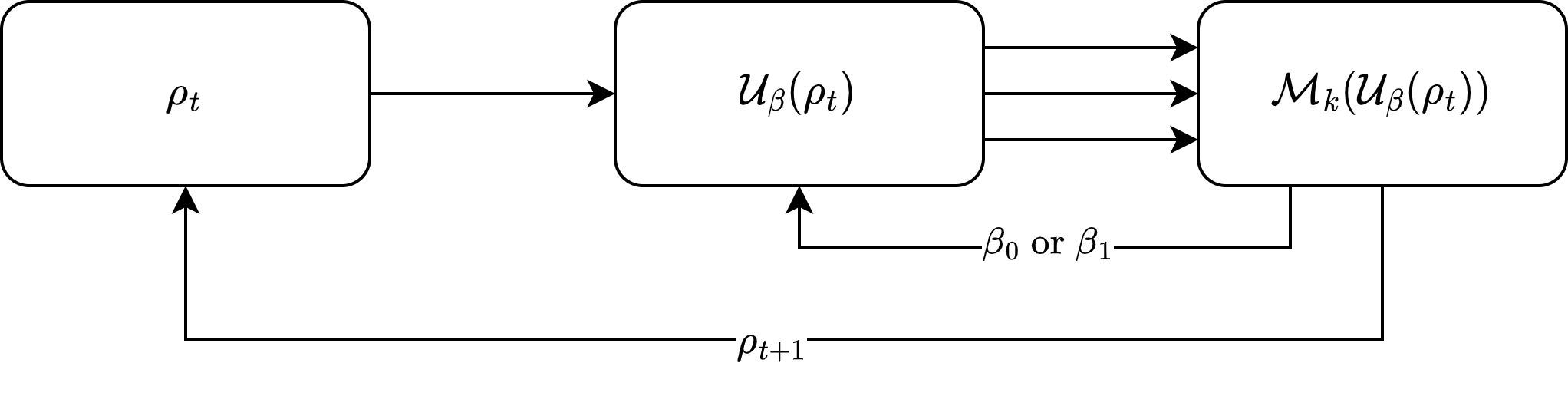}
    \caption{Basic Controller block scheme}
    \label{fig:eDeterministic_controller}
\end{figure}

The values of the two parameters that maximize the previous figure of merit are $\beta_0 = \beta_1 = 1$.

\subsection{Model-Based learning scenario}
In the Model-Based scenario (MBs), we consider a quantum system with a known initial state represented by the density operator $\bar{\rho_0}$. In this scenario, we assume the agent has access the system's state at time $t$, the controls $\beta_t$, and the measurement set $\{M_l^\epsilon\}$. During the training phase, the agent is training on the evolution and measurements statistics associated to the \emph{nominal dynamics} \eqref{eq:nominal}. Indeed in this phase the noise operator, denoted by $\mathcal{N}_{\alpha}$, is not included in the model $(\alpha = 0)$, and the
 distribution of the measurement outcomes it the one provided in equation \eqref{eq: Nominal_state_dist}.

In this phase, the agent receives the current state of the quantum system $\bar{\rho_t}$ as input at each time step $t$ and outputs the control parameter $\beta$. 
The agent learns to make decisions about $\beta$ based on the observed state and the ultimate goal, which consists in the maximization of the fidelity function as figure of merit. 

During the validation phase, noise is injected in the system ($\mathcal{N}_\alpha$ with $\alpha \neq 0$), so that the dynamic of the system follow equation \ref{eq:dynamics}. The agent has to choose the control parameter $\beta$ at every timestep $t$, but this time the dynamics of the gym environment follow the \emph{filtered equation} \eqref{eq:filtered}.
\begin{figure}[h]
    \centering
    \includegraphics[width = 8.6cm]{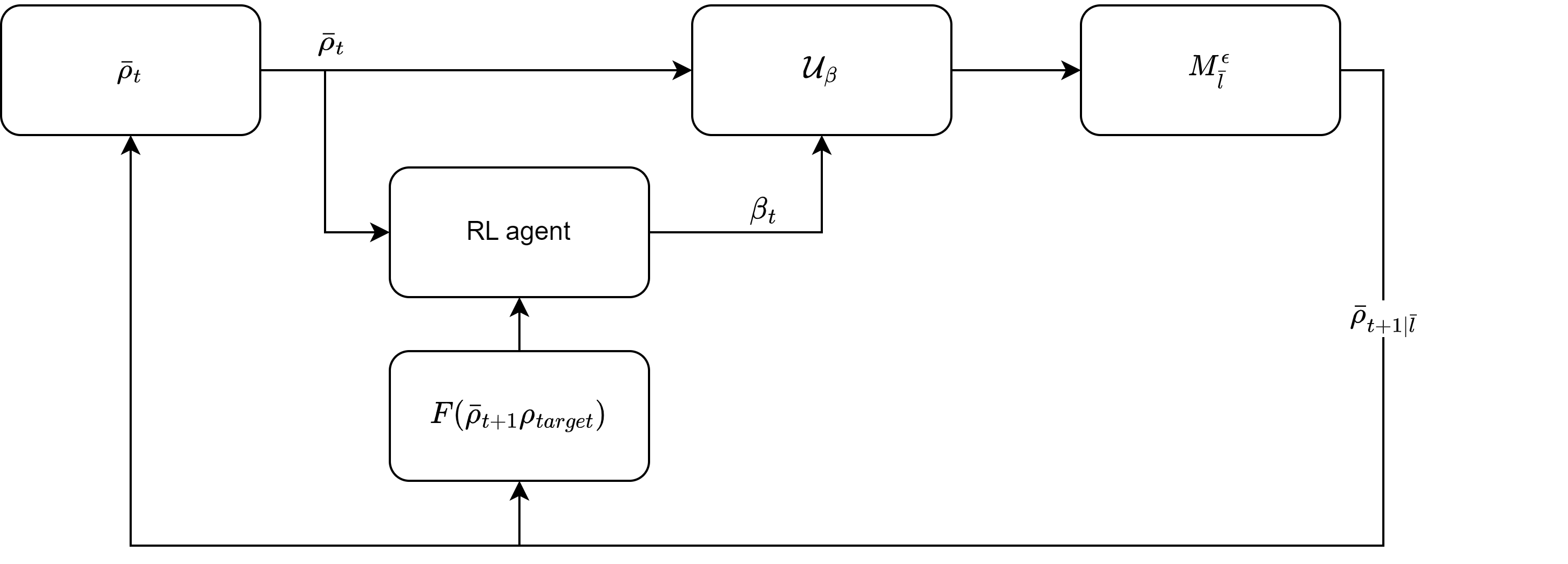}
    \caption{Training scheme of the Model-Based scenario}
    \label{fig:Model_Based}
\end{figure}

\subsection{Data-Based learning scenario}
In the Data-Based scenario (DBs), we consider a quantum system with a known initial state represented by the density operator $\hat{\rho_0}$. As for the one before, the agent is aware of the system's state at time $t$, the control parameter $\beta$, and the measurement set $M$. During the training phase the dynamics of the gym environment where the agent is trained, it is represented by the \emph{filtered state}. When we get a measurement outcome, that in this case it is sampled from the real distribution (eq \ref{eq: Filtered_distr}), we compute the current state with the best estimation possible, which is given from the following filtering equation:

\begin{equation}
     \hat{\rho}_{t+1} =\frac{M_{\hat{l}_t}^\epsilon U_\beta \hat{\rho_t} U_\beta^\dagger M_{\hat{l}_t}^{\epsilon^\dagger}}{tr(\cdot)}.
\end{equation}

In the training phase we fed the agent with the estimate of the current state $\rho_t$, and we collect as output the control parameter $\beta(t)$. Also in this scenario the reward function the agent tries to maximize is the fidelity, which is computed as in equation (\ref{eq: Fidelity})

During the validation phase we keep the same dynamics as in the training phase. Also this time the agent has to select at every timestep $t$ the best control parameter $\beta$ to stabilize the state.

\begin{figure}[h]
    \centering
    \includegraphics[width = 8.6cm]{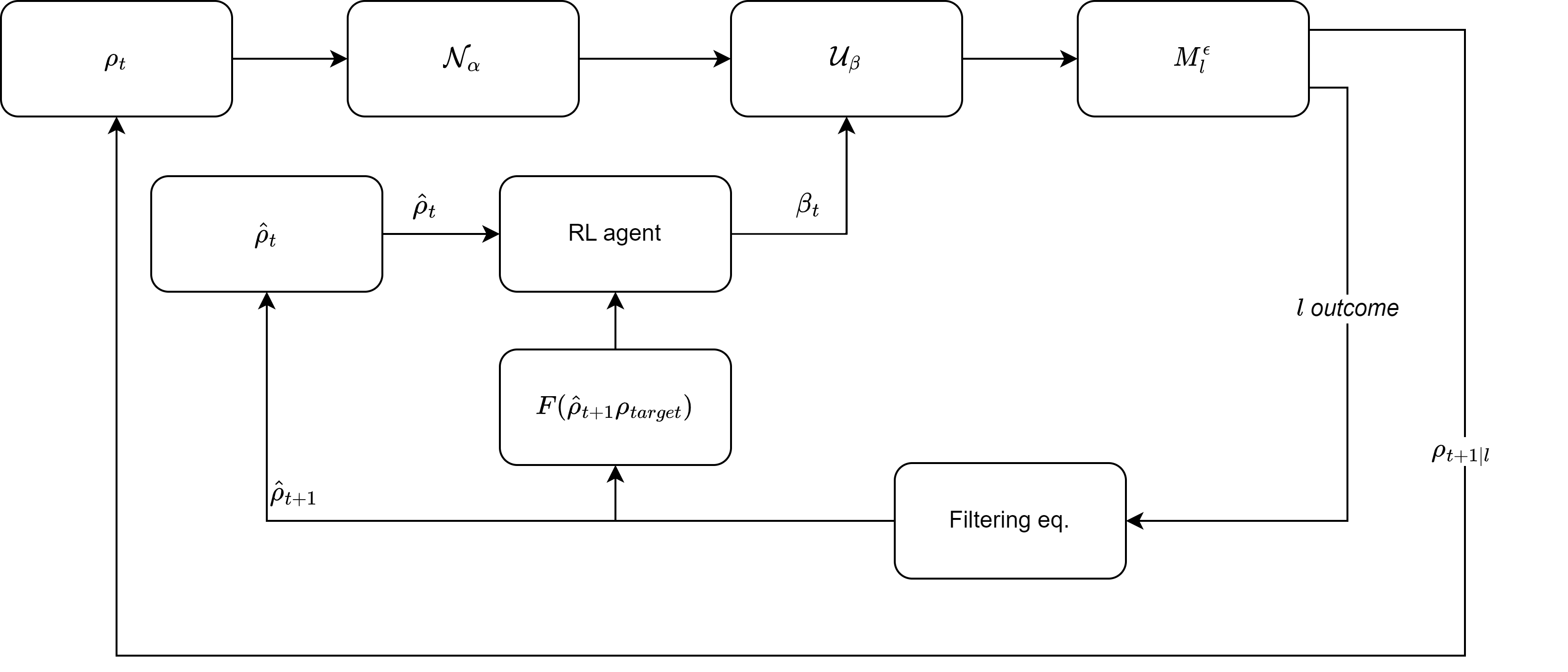}
    \caption{Training scheme of the Data-Based scenario}
    \label{fig:Data_Based}
\end{figure}

\subsection{Model-Free Scenario: Quantum Observable Markov Decision Process}
In the model-free scenario we train the controller without any reference  model, quantum or classical. The approach will be called Quantum Observable Markov Decision Processes (QOMDPs), as it is inspired by  the works \cite{DBLP:journals/corr/BarryBA14,Sivak_2022}. We consider a quantum system characterized by an initial state $\rho_0$ that evolves over time according to its dynamics, which depends on a noise parameter $\alpha$ (as specified in Equation \ref{eq:dynamics}). In this scenario, the distribution of measurement outcomes follows the true distribution of outcomes, similar to the DBs approach.

The agent in this QOMDP framework does not receive the state of the quantum system $\rho_t$ or any estimation thereof, as is the cases of MBs or DBs. It receives only two pieces of information at each time step $t$: the outcome of the last measurement denoted as $l_{t-1}$ and the previous control action denoted as $\beta_{t-1}$.

At each time step $t$, the agent has to make two decisions: selecting the control action $\beta_t$ and specifying the value of the \emph{stop action}. If the stop action equals 1, the episode ends; otherwise, if it takes the value 0, the episode continues. At the initial time step $t = 0$, no action is performed, i.e., $\beta = 0$ and $stop = 0$. This leads to a unitary evolution represented by the identity operator $U(0) = I_{n\times n}$. After this, the first state provided to the agent is always in the form of the column vector $[l_{t=0}, 0]^T$.

To set up the training environment, we introduce a new set of measurements called the \emph{last observation} denoted as $l_{last}$. This observation is obtained by measuring the quantum system using a projective set of measurement operators ${M^{end}_0, M^{end}_1, M^{end}_2 }$ as follows:

\begin{equation}
    \begin{aligned}
        M^{end}_0 = \begin{bmatrix}
        1 & 0 & 0\\
        0 & 0 & 0\\
        0 & 0 & 0
        \end{bmatrix},\\
        M^{end}_1 = \begin{bmatrix}
        0 & 0 & 0\\
        0 & 1 & 0\\
        0 & 0 & 0
        \end{bmatrix},\\
        M^{end}_2 = \begin{bmatrix}
        0 & 0 & 0\\
        0 & 0 & 0\\
        0 & 0 & 1
        \end{bmatrix}.
    \end{aligned}
\end{equation}

Once the last observation $l_{last}$ is measured, we can compute the reward function as follows:

\begin{equation}
R(stop, l_{last}, done) =
\begin{cases}
0, & \text{if }stop = 0 \: done = \text{False}\\
-1, & \text{if }stop = 0 \: done = \text{True}\\
r_{fun}(l_{last}), & \text{if } stop = 1
\end{cases},
\end{equation}

Where $r_{fun}$ is defined as:

\begin{equation}
r_{fun}(l_{last}) =
\begin{cases}
+1, & \text{if } l_{last} = l_{target}\\
-1, & \text{otherwise}
\end{cases},
\end{equation}

In this expression, $l_{target}$ represents a target observation value for the quantum state we want to reach.

\begin{figure}[h]
    \centering
    \includegraphics[width = 8.6cm]{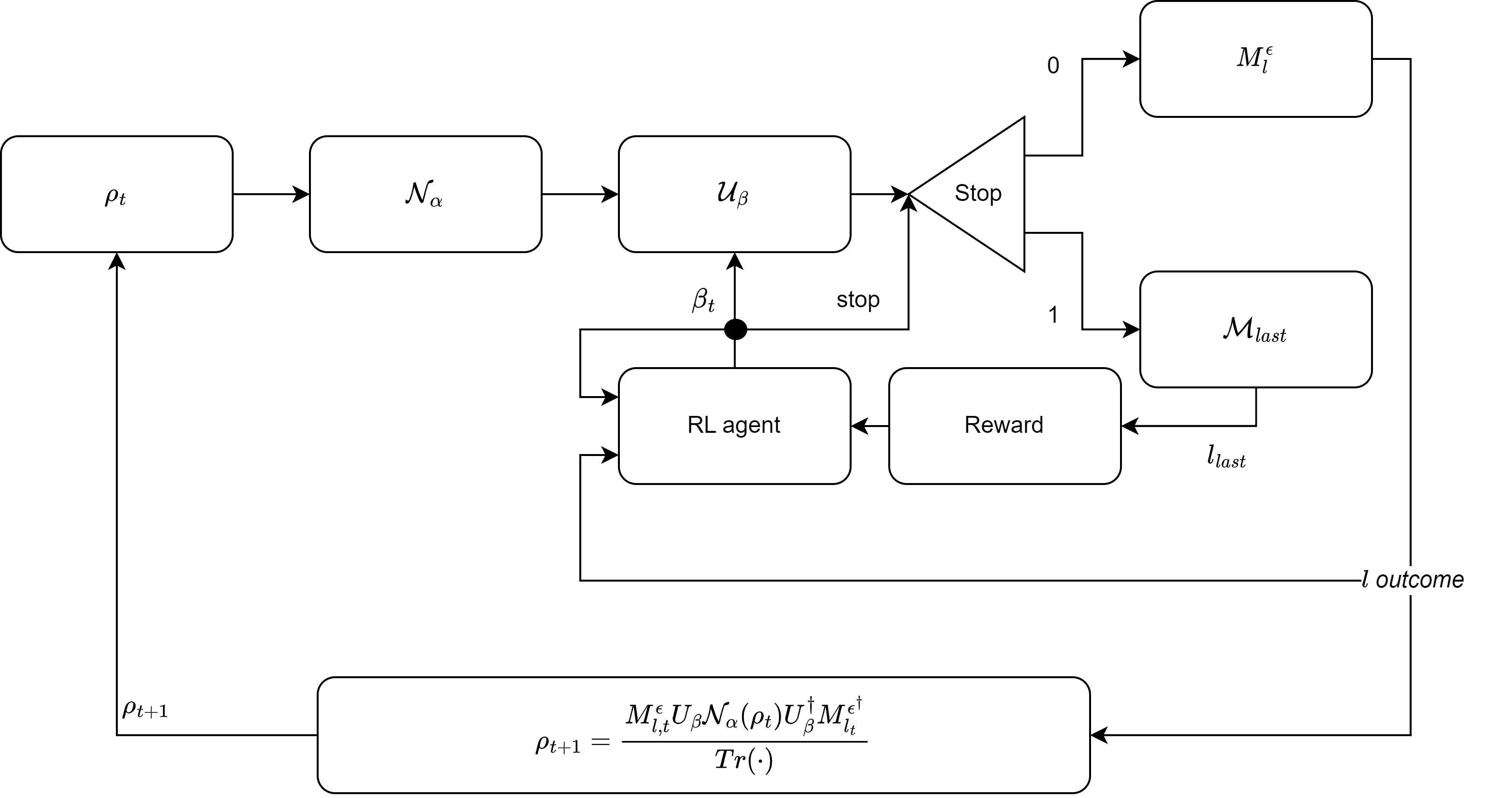}
    \caption{Training scheme of the QOMDP scenario}
    \label{fig:QOMDP}
\end{figure}

During the validation phase, the quantum system's dynamics remain unchanged. The reward function is no longer utilized or considered, since it is part of the learning process. The primary focus is on evaluating the agent's performance and decision-making without incorporating the reward function. The agent's actions are assessed solely based on their impact on the quantum system's evolution and fidelity w.r.t. the target state.

%%%%%%%%%

\section{Numerical Analysis}\label{sec:numerical} 
{\bf Experiments setup.} In this section of the paper, we report on the results of the simulations\footnote{The primary research efforts were conducted at the University of Padova, while certain simulations concerning DBs and MBs were performed at FZJ.} in order to assess the performance of various RL models with respect to variation in the measurement accuracy and the noise level. We summarize in the following table the main peculiarities of each model.
\begin{table}[h]
    \centering
    \begin{tabular}{|c|c|c|c|c|} \hline 
        Model & Trainig & Train $\alpha$ & Train $\epsilon$ & Validation\\ \hline
        MBs & Nominal Dyn. & 0 & every $\epsilon$ & Filtered Dyn.\\ \hline 
        DBs & Filtered Dyn. & every $\alpha$ & every $\epsilon$ & Filtered Dyn.\\ \hline 
        QOMDP & Nominal Data & 0 & every $\epsilon$ & Real Data\\ \hline
    \end{tabular}
    \caption{RL Models Summary}
    \label{tab:my_label}
\end{table}
Our comparative analysis focused on evaluating the fidelity of the state $\rho(t)$, which was generated by applying the true dynamics while accounting for noise and utilizing control parameters determined by the RL agent or by the basic controller. 
We considered the three types of noise that we had previously introduced: the random permutation noise, the depolarizing channel, and the amplitude damping channel. 

We trained and subsequently tested the various RL models for all possible configurations of the following parameters \footnote{The code is available at \url{https://github.com/ManuelGuatto/RL_4_Robust_QC.git} }.
\begin{table}[h]
    \centering
    \begin{tabular}{|c|c|l|c|c|c|c|c|c|c|c|l|} \hline 
         Parameter&   \multicolumn{11}{|c|}{Values}\\ \hline 
         $\alpha$&   0&0.1&  0.2&  0.3&  0.4&  0.5&  0.6&  0.7&  0.8& 0.9 &1\\ \hline 
         $\epsilon$&   0.1&0.15&  0.175&  0.2&  0.25&  0.3&  \multicolumn{5}{|c|}{}\\ \hline
    \end{tabular}
    \caption{Training and Test Parameters}
    \label{tab:train_test_params}
\end{table}

To ensure a robust evaluation of our simulations, we adopted a standardized approach. Specifically, for each simulation, we gathered a dataset consisting of 1000 samples. Subsequently, we computed both the mean and the standard deviation from this dataset, visually represented by lines and shaded regions. respectively, in the plots.

It is noteworthy that our modeling approaches, MBs and DBS, were trained using the PPO algorithm. For these methods, we employed a conventional Multi-Layer Perceptron (MLP) network architecture. In contrast, the QOMDP approach underwent training using PPO in conjunction with a Long Short-Term Memory (LSTM) network. This choice was made to enable the tracking of previous measurement outcomes.

\begin{figure*}[t]
   { \centering
        \includegraphics[width=1\textwidth]{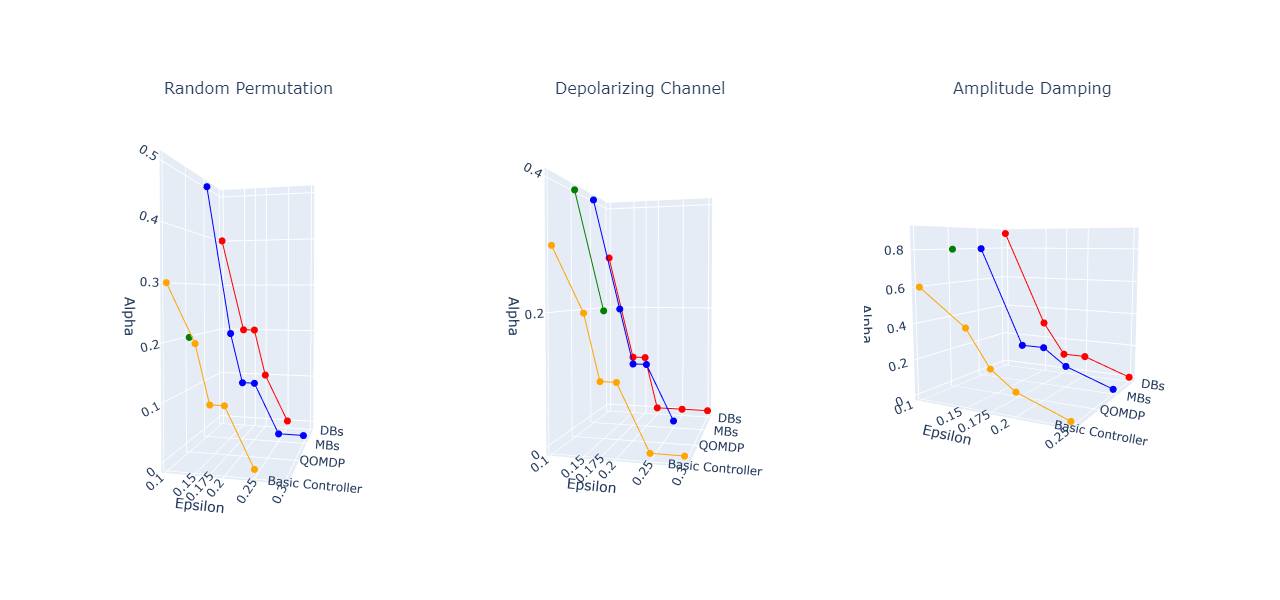}}
        \caption{In this figure we represent the highest value of noise intensity $\alpha$ that can be handled by each controller while maintaining a fidelity of at least $0.9$. Missing points in the curve indicate that the target fidelity cannot be guaranteed.}
        \label{fig:90_Fidelity_Comparison}
\end{figure*}

{\bf Results.} In the first plot (Figure \ref{fig:90_Fidelity_Comparison}) we display the highest intensity of noise that an agent can handle while maintaining a final fidelity of at least $0.9$, for each $\epsilon$ . 
Some observations follows from the comparison of figure \ref{fig:90_Fidelity_Comparison}. Firstly, when dealing with precise measurements (i.e., with $\epsilon = 0.1$), it is evident that the RL agents exhibit superior performance compared to the Basic controller.

Notably, when confronted with random permutation noise, both the DBs and MBs agents demonstrate their robustness by effectively managing the situation up to $\alpha = 0.5$. Similarly, when faced with the challenges of introducing the depolarizing channel noise at $\epsilon = 0.1$, the MBs scenario outperforms other approaches, extending its capability to handle noise levels up to $\alpha = 0.4$. 
For the amplitude damping channel, it is evident across all considered scenarios that controllers exhibit a higher noise threshold compared to the other two noise models - see Figure \ref{fig:90_Fidelity_Comparison} at least for high measurement accuracy ($\epsilon\leq 0.15$). Notably, the DBs agent demonstrates superior performance, achieving target fidelity value with noise intensity $\alpha=0.9$. Following closely are both the MBs and QOMDP approaches, attaining similar fidelity values till $\alpha=0.8$. The deterministic approach remains limited, achieving target fidelity for $\alpha=0.6$. For $\epsilon>0.1$,  MBs, DBs, and the deterministic controller exhibit comparable performances. On the other hand, the performance of the QOMDP approach degrades quickly, and does not reach target fidelity $\epsilon>0.1.$

It is worth observing that in general the QOMDP agent displays commendable performance as long as measurements are informative. This outcome can be attributed to the reliance of the QOMDP agent on the quality this latter, which plays a pivotal role in determining when to execute control actions effectively. On the other hand, it is not able to guarantee target fidelity as soon as the measurement quality decreases.
The simulations also confirm the presence a {\em trade-off between measurement accuracy and the level of noise} that all the controllers can effectively cope with.

To further elucidate this phenomenon, we present two paradigmatic plots illustrating the performance of the various methods at various noise levels when $\epsilon = 0.175$.

\begin{figure*}
    \begin{subfigure}[t]{0.5\textwidth}
        \centering
        \includegraphics[width=0.9\textwidth]{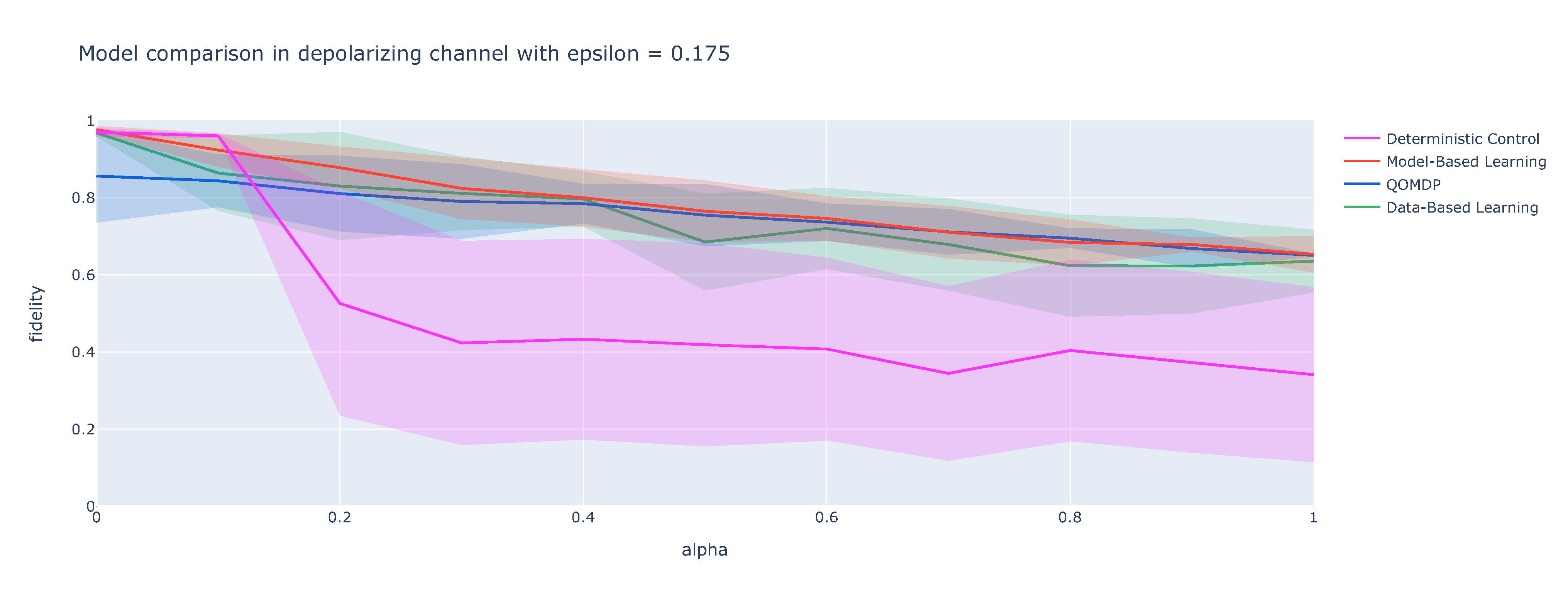}
        \caption{Depolarizing channel}
        \label{fig:Depolarizing}
    \end{subfigure}%
    \begin{subfigure}[t]{0.5\textwidth}
        \centering
        \includegraphics[width=0.9\textwidth]{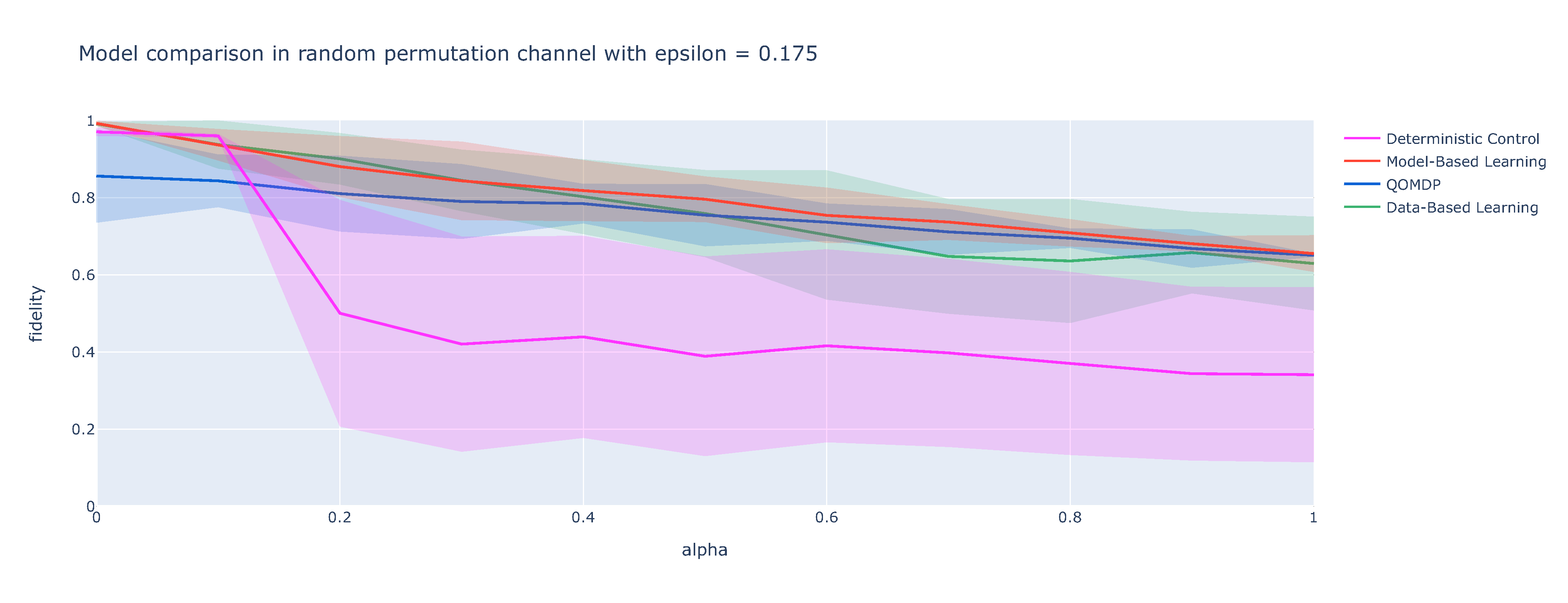}
        \caption{Random permutation channel}
        \label{fig:Random_Permutation}
    \end{subfigure}
            
    \caption{Simulation results for $\epsilon = 0.175$}
    \label{fig:results_175}
\end{figure*}
These two plots offer some critical insights: (i) while there is an initial resemblance in behavior between the Basic Controller and the MBs and DBs agents when confronted with low noise intensity, as the noise intensity escalates the performance of the basic controller drops; (ii) the performance of the QOMPD is lower than the other RL methods.

It is important to note that we do not contend that the fidelity achieved by the RL agents is necessarily optimal. Instead, our contention is that RL presents a more robust control approach compared to the Basic Controller. In other words, it excels in maintaining control performance as noise levels increase.

Last, it is possible to note that the number of necessary steps in order to achieve a fidelity greater or equal than $0.9$ in general is lower for the RL agents compared to the basic controller. This result is summarized in the figure \ref{fig:Timesteps_Comparison}.
Summarizing the findings of the extended simulations we ran, we can conclude that:
\begin{itemize}
\item For low noise intensity, the basic controller and the RL controllers have similar performance; 
\item It is enough however to raise the measurement inaccuracy $\epsilon$ to $0.1$ to notice a significant advantage of the RL controllers in terms of the amount of noise they can withstand while maintaining the fidelity threshold;

\item The model-based controller has performance that compete with the data-based one; 
\item The trade-off between measurement accuracy and noise is evident, and becomes less taxing for RL-based controllers. 

\item The RL controller tends to converge to high fidelity in less steps than the basic controller.
\begin{figure}[ht]
    \centering
    \includegraphics[width = 5.6cm]{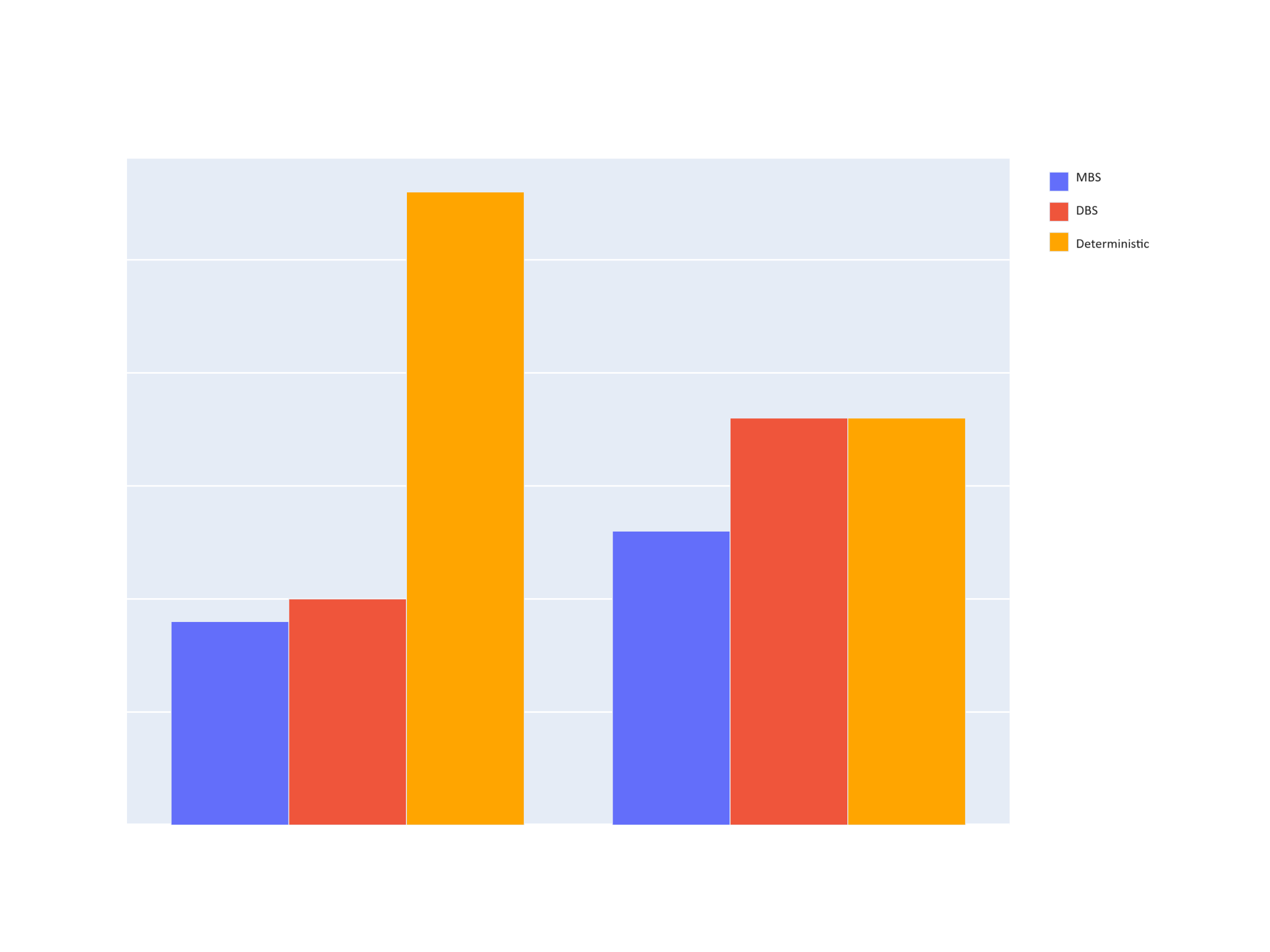}
    \caption{Examples of necessary number of timesteps to achieve fidelity $\geq 0.9$}
    \label{fig:Timesteps_Comparison}
\end{figure}

\end{itemize}

\section{Conclusions}\label{sec:conclusions} 
In this study, we conducted an assessment of the performance of various reinforcement learning (RL) models in controlling quantum systems subject to noise and uncertainties towards a state preparation task. Our performance analysis analysis focused on the fidelity of a quantum state $\rho(t)$ with respect to a target pure state, which was generated by applying true dynamics while accounting for noise and utilizing control parameters determined by RL agents or a basic controller. 

The study investigated a range of RL models, including Model-Based scenario (MBs), Data-Based scenario (DBs), and Quantum Observable Markov Decision Processes (QOMDP). To empower these models, we utilized the Proximal Policy Optimization (PPO) algorithm and distinct network architectures.

In the Model-Based Scenario (MBs) and Data-Based Scenario (DBs) the agents harnessed the PPO algorithm, paired with a traditional Multi-Layer Perceptron (MLP) network architecture. Their training phases were based on data obtained from the nominal dynamics (as defined in Equation \ref{eq:nominal}) for the MBs, and real data fed to the filtering dynamics (as defined in Equation \ref{eq:filtered}) equations for the DBs. 

In contrast to MBs and DBs, the Quantum Observable Markov Decision Processes (QOMDP) agent employed PPO in conjunction with a Long Short-Term Memory (LSTM) network architecture. This approach allowed the QOMDP agent to track and incorporate previous measurement outcomes into its decision-making process. Importantly, the QOMDP agent did not construct an updated system state; instead, its actions were solely informed by measurement outcomes.

We considered a testbed system on which running the simulations in order to compare the behaviour of the different control scenarios. We tested the performance of each model w.r.t. both the depolarizing channel and a random permutation channel accounting for different noise intensity and measurements inaccuracy parameters (table \ref{tab:train_test_params}).

To establish a benchmark for the aforementioned models, we introduced a reference control strategy, referred to as the basic controller. This basic controller is designed to obtain perfect state preparation when both uncertainty in measurements and noise intensity parameters, $\epsilon$ and $\alpha$, are set to zero , maximizing the probability of transitioning from the quantum states $\ket{0}\bra{0}$ and $\ket{1}\bra{1}$ to the target state $\ket{2}\bra{2}$.

The numerical analysis we performed analyzed the performance of various controllers in the presence of noise and measurement inaccuracy:

In scenarios with low noise intensity, both the basic controller and RL controllers demonstrate comparable performance. This implies that traditional control methods can be as effective as reinforcement learning-based controllers in relatively noise-free environments.

However, when we increase the measurement inaccuracy to a modest value of $\epsilon = 0.1$, the RL controllers exhibit a clear advantage. They excel in withstanding noise while still maintaining the desired fidelity threshold (greater or equal than $0.9$). This showcases the robustness and adaptability of RL-based control in challenging conditions.

Our results also highlight a trade-off between measurement accuracy and noise tolerance for all methods, but it imposes less restrictive constraints for RL-based controllers. They appear to be better suited to navigate this delicate balance.

Moreover, RL controllers require fewer steps to reach a high level of fidelity compared to the basic controller. This suggests that RL methods can offer faster and more efficient control solutions in certain scenarios.

In summary, our study underscores the adaptability and resilience of reinforcement learning controllers in the face of noise and measurement inaccuracy. While traditional control methods are effective in low noise conditions, RL controllers show a clear advantage when confronted with more challenging and uncertain environments. The choice between these approaches should be made based on the specific requirements and conditions of the task at hand.

\nocite{*}

\bibliography{References}% Produces the bibliography via BibTeX.

\section*{Appendix}

\textbf{Add the specs of all the other parameters}

\subsection*{Network architectures and training parameters}

This appendix provides a detailed description of the network architecture and hyperparameters employed in the experiments. An understanding of these details is crucial for replicating and extending the presented work.\\

\textbf{MBs and DBs scenarios.} The Actor-Critic policy networks employed in the MBs and DBs scenarios consist of three main components:

\begin{itemize}
    \item \textbf{Flatten Extractor:} This component flattens the input state vector, converting it into a one-dimensional representation.

    \item \textbf{Policy and Value Function Extractors:} These extractors further process the flattened state representation using two separate MLP architectures, one for policy estimation and the other for value function approximation. Each MLP consists of three hidden layers with 64 nodes each, using Tanh activation functions.

    \item \textbf{Action and Value Nets:} These nets receive the output of the respective extractors and generate the policy distribution and value function estimate, respectively. The action net is a single-layer linear transformation with one output node, representing the predicted action probabilities. The value net also utilizes a single-layer linear transformation to output a scalar representing the predicted state value.
\end{itemize}

The trainings of the networks for these two scenarios were conducted with the following parameters:

\begin{itemize}
    \item \textbf{Batch Size:} The batch size for each training update was set to 512. This value represents the number of state-action pairs used to update the network parameters.

    \item \textbf{Number of Steps per Update:} For each training update, the agent interacts with the environment for a total of 512 steps. This corresponds to the number of state-action pairs sampled before updating the policy and value function parameters.

    \item \textbf{Learning Rate:} The learning rate for the optimizer was set to 1e-4. This value controls the step size during gradient descent updates, ensuring a balance between exploration and exploitation.
\end{itemize}

For all the other parameters we used the default stable-baselines3 settings.

\textbf{QOMDP scenario}. The LSTM policy network employed in this scenario consists of four main components:

\begin{itemize}
    \item \textbf{Flatten Extractor:} This component flattens the input state vector, converting it into a one-dimensional representation.

    \item \textbf{Policy and Value Function Extractors:} These extractors further process the flattened state representation using two separate MLP architectures, one for policy estimation and the other for value function approximation. Each MLP consists of three hidden layers with 64 nodes each, using Tanh activation functions.

    \item \textbf{Action and Value Nets:} These nets receive the output of the respective extractors and generate the policy distribution and value function estimate, respectively. The action net is a single-layer linear transformation with two output nodes, representing the predicted action probabilities for the two possible actions. The value net also utilizes a single-layer linear transformation to output a scalar representing the predicted state value.

    \item \textbf{Recurrent Architecture:} To capture temporal dependencies in the environment, the network utilizes two LSTM units, one for policy and one for value function estimation. These LSTM units process the sequences of state and action inputs, enabling the network to learn long-range dependencies and adapt its behavior accordingly.
\end{itemize}

The training of the Recurrent Actor-Critic policy network was conducted with the following parameters:

\begin{itemize}
    \item \textbf{Batch Size:} The batch size for each training update was set to 512. This value represents the number of state-action pairs used to update the network parameters.

    \item \textbf{Number of Steps per Update:} For each training update, the agent interacts with the environment for a total of 512 steps. This corresponds to the number of state-action pairs sampled before updating the policy and value function parameters.

    \item \textbf{Learning Rate:} The learning rate for the optimizer was set to 3e-4. This value controls the step size during gradient descent updates, ensuring a balance between exploration and exploitation.
\end{itemize}

For all the other parameters we used the default stable-baselines3 settings.

\end{document}